\documentclass[11pt]{article}

\usepackage[top=50mm, bottom=50mm, left=50mm, right=50mm]{geometry}

\usepackage{amsmath}
\usepackage{amssymb}
\usepackage{amsfonts}
\usepackage{mathtools}
\usepackage{mathrsfs}
\usepackage[scr = dutchcal]{mathalpha}
\usepackage{graphicx}
\usepackage{graphics}
\usepackage{float}
\usepackage{color}
\usepackage{fullpage}
\usepackage[normalem]{ulem} 
\usepackage{makeidx}
\usepackage{xspace}
\usepackage{authblk}
\usepackage{datetime}
\usepackage{microtype}
\usepackage{xpatch}
\usepackage{xstring}
\usepackage{fancyhdr}
\usepackage{mathrsfs}
\usepackage{esint}
\usepackage{upgreek}
\usepackage{siunitx}
\usepackage{chemformula}
\usepackage{tabularx, multirow}
\usepackage{threeparttable}
\usepackage[]{booktabs} 
\usepackage{float}
\usepackage{graphicx,xcolor}
\usepackage[colorlinks=false]{hyperref}
\usepackage[noabbrev]{cleveref}
\usepackage[font = footnotesize, labelfont=bf]{caption}

\fancypagestyle{fpg}{%
	\lfoot{\footnotesize Applied Sciences \textbf{2021}}%
	\cfoot{\normalsize\thepage}
	\rfoot{\footnotesize \href{}{}}%
}

\makeindex

\begin{document}

\title{\Large\textbf{Pressure-driven Nitrogen Flow in Divergent Microchannels with Isothermal Walls}}
 
	\author[1,$\dagger$]{Amin Ebrahimi}
	\author[2]{Vahid Shahabi}
	\author[3]{Ehsan Roohi}
	\affil[1]{\small\textit{Department of Materials Science and Engineering, Faculty of Mechanical, Maritime and Materials~Engineering, Delft University of Technology, Mekelweg 2, 2628~CD~Delft, The~Netherlands}}
	\affil[2]{\small\textit{Department of Mechanical Engineering, Faculty of Engineering, Ferdowsi University of Mashhad, Mashhad, P.O.~Box~91775-1111, Khorasan Razavi, Iran}}
	\affil[3]{\small\textit{State Key Laboratory for Strength and Vibration of Mechanical Structures, International Center for Applied Mechanics, School of Aerospace Engineering, Xi'an Jiaotong University, Xi'an, China}}
	\affil[$\dagger$]{\textit{Corresponding author, Email: A.Ebrahimi@tudelft.nl}}
	
\date{}
\maketitle
\thispagestyle{fpg}

\begin{abstract}
		Gas flow and heat transfer in confined geometries at micro and nano scales differ considerably from those at macro-scales, mainly due to nonequilibrium effects such as velocity slip and temperature jump. The~nonequilibrium effects enhance with a~decrease in the~characteristic length-scale of the~fluid flow or the~gas density, leading to the~failure of the~standard Navier-Stokes-Fourier~(NSF) equations in predicting thermal and fluid flow fields. The~direct simulation Monte-Carlo~(DSMC) method is employed in the~present work to investigate pressure-driven nitrogen flow in divergent microchannels with various divergence angles and isothermal walls. The~thermal fields obtained from numerical simulations are analysed for different inlet-to-outlet pressure ratios ($1.5 \leq \Pi \leq 2.5$), tangential momentum accommodation coefficients and Knudsen numbers ($0.05 \leq \mathrm{Kn} \leq 12.5$), covering slip to free-molecular rarefaction regimes. The~thermal field in the~microchannel is predicted, heat-lines are visualised, and the~physics of heat transfer in the~microchannel is discussed. Due to the~rarefaction effects, the~direction of heat flow is largely opposite to that of the~mass flow. However, the~interplay between thermal and pressure gradients, which are affected by geometrical configurations of the~microchannel and applied boundary conditions, determines the~net heat flow direction. Additionally, the~occurrence of thermal separation and cold-to-hot heat transfer (also known as anti-Fourier heat transfer) in divergent microchannels is explained.
\end{abstract}

\bigskip
\newpage

\section{Introduction}
\label{sec:intro}

Gas flow in micro- and Nano-channels with non-uniform cross-sections offers opportunities to develop small devices with novel applications (see for example,~\cite{W_rger_2011,Gebhard_1996,Jiang_1998,Duryodhan_2016,Yang_2016,Bordbar_2020}). Understanding heat and fluid flow in microchannels is essential to engineer novel micro-electromechanical systems (MEMS)~\cite{Agrawal_2011,Ebrahimi_2015,Ebrahimi_2016b}. However, this is a~challenging task since the~rate of collisions between the~gas molecules and solid walls reduces with decreasing the~characteristic length scale of the~fluid flow or the~gas density, affecting the~random movement of molecules, which is commonly known as nonequilibrium effects~\cite{Ferziger_1973,Johnson_2016}. Nonequilibrium molecular transport processes cause velocity slip and temperature jump at walls and influence thermal and fluid flow fields~\cite{Gad-el-Hak_2005,Kirby_2010}. The~Knudsen number~(Kn), which is the~ratio of the~molecular mean free path~$\lambda$ to a~characteristic length scale~$\mathscr{L}$, is often employed as an~indicator of deviation from the~equilibrium condition. The~higher the~Knudsen number, the~higher the~deviation from the~equilibrium condition. Four different rarefaction regimes have been defined based on the~Knudsen number~\cite{Roy_2003}, and are commonly named as free molecular ($\mathrm{Kn} > 10$), transition ($10^{-1} \leq \mathrm{Kn} \leq 10$), slip ($10^{-3} \leq \mathrm{Kn} \leq 10^{-1}$) and continuum ($\mathrm{Kn} \leq 10^{-3}$) regimes.

Micro-devices usually operate in the~slip and transition rarefaction regimes at the~standard pressure and temperature~\cite{Gad-el-Hak_2005}. It is widely acknowledged that the~standard Navier-Stokes-Fourier~(NSF) equations fail to predict thermal and fluid flow fields accurately when the~nonequilibrium effects are significant~\cite{Karniadakis_2005,Shen_2005,Kara_2017}. The~Boltzmann equation governs fluid flow at micro-scales in the~entire range of Knudsen number~\cite{Bird_1994}. However, direct numerical solution of the~Boltzmann equation is generally expensive with available computational capabilities and are limited to simple applications because of the~high dimensionality of the~Boltzmann equation and the~complexity of the~collision integral. Moreover, the~application of the~Boltzmann equation is limited to dilute gasses due to the~assumption of binary intermolecular collisions~\cite{Jakobsen_2008}. The~direct simulation Monte-Carlo~(DSMC) method is a~numerical technique that has been developed based on the~kinetic theory to approximate the~solution of the~Boltzmann equation~\cite{Bird_1994}.

Zheng~\textit{et al.}~\cite{Zheng_2002} compared the~numerical predictions obtained from DSMC simulations with the~solutions of the~NSF equations for a~Poiseuille gas flow between two parallel plates, and reported quantitative and qualitative differences between the~predictions. Deviations between numerical predictions obtained using the~DSMC method and the~NSF equations increases with increasing the~Knudsen number~\cite{Zheng_2002,ARKILIC_2001,Arlemark_2008,Qazi_Zade_2012}. Using the~regularised 13-moment equations, Torrilhon and Struchtrup~\cite{Torrilhon_2009} achieved a~better agreement with the~DSMC results up to $\mathrm{Kn} \approx 1$ in comparison with the~NSF-based solutions. Varoutis~\textit{et al.}~\cite{Varoutis_2012} studied Poiseuille gas flow in channels with finite lengths using the~DSMC method, and stated that the~nonequilibrium effects are the~most significant factor affecting rarefied gas flow in channels.

The~majority of published literature on subsonic internal gas flows at micro-scales are limited primarily to flow passages with uniform cross-sections~\cite{Kandlikar_2013,Fang_2002,Liou_2001}. Moreover, previous studies have mostly focused on characterising pressure loss, mass flow rate, and velocity field~\cite{Zhang_2017}. There is a~crucial lack of detailed investigations concerning the~thermal field analysis of gas flow in micro- and Nano-channels with non-uniform cross-sections in which nonequilibrium effects play an~important role~\cite{Hemadri_2018}. Using the~NSF equations with the~first-order slip boundary condition, Varade \textit{et al.} calculated temperature variations in divergent~\cite{Varade_2015} and convergent~\cite{Varade_2015_2} microchannels for both continuum and slip flow regimes. Using the~same approach, Hemadri~\textit{et al.}~\cite{Hemadri_2016,Hemadri_2017} calculated temperature distribution in convergent microchannels in the~slip and early-transition rarefaction regimes. Mili{\'{c}}ev and Stevanovi{\'{c}}~\cite{Milicev_2020} proposed an~analytical solution to the~one-dimensional NSF equations to describe steady-state pressure-driven isothermal gas flow in microchannels with variable cross-sections. Ohwada~\textit{et al.}~\cite{Ohwada_1989} stated that heat and mass flow are in opposite directions for Poiseuille flow between two parallel plates for $\mathrm{Kn} > 10^{-1}$ while they are in the~same direction adjacent to the~walls at low Knudsen numbers. It should be noted that only the~tangential heat flow profiles were considered in their study while both normal and tangential heat fluxes contribute to the~net heat flow. John~\textit{et al.}~\cite{John_2013} numerically studied the~influence of pressure ratio and surface accommodation coefficient on the~thermal field of a~Poiseuille gas flow between two parallel plates. The~DSMC method has been recently employed to investigate temperature variations in divergent microchannels~\cite{Ebrahimi_2016,Ebrahimi_2017,Guerrieri_2016}. Characterising the~thermal field in microchannels with non-uniform cross-sections is emerging for micro- and Nano-system engineering and further investigations are essential.

Despite the~extensive interest in studying fluid flow at the~micro- and Nano-scales, numerical investigations of gas flow under nonequilibrium conditions are deficient in thermal analysis. Previous studies on thermal field analysis in microfluidics systems are mostly limited to the~channels with uniform cross-sections (see for instance,~\cite{Kannan_2020,Taassob_2018,John_2013,Shah_2018,Balaj_2017,Gavasane_2017,Hong_2012,Roohi_2009,Hong_2007}). Therefore, there is an~indispensable need for detailed investigations on heat and fluid flow at the~micro- and Nano-scales under nonequilibrium conditions. The~goal of the~present study is to enhance our understanding of heat and fluid flow in divergent microchannels, which are applicable in micro-devices such as micro-pumps, micro-actuators and micro-thrusters. Numerical simulations based on the~DSMC method are performed to explore the~effects of divergence angle, inlet-to-outlet pressure ratio, rarefaction, and tangential momentum accommodation coefficient on thermal and fluid flow fields. Heat and fluid flow under nonequilibrium conditions are considerably different from its equilibrium counterpart, providing the~motivation to conduct the~present study.

\section{Problem description}
\label{sec:problem}

\begin{figure}[H] 
	\centering
	\includegraphics[width=0.8\linewidth]{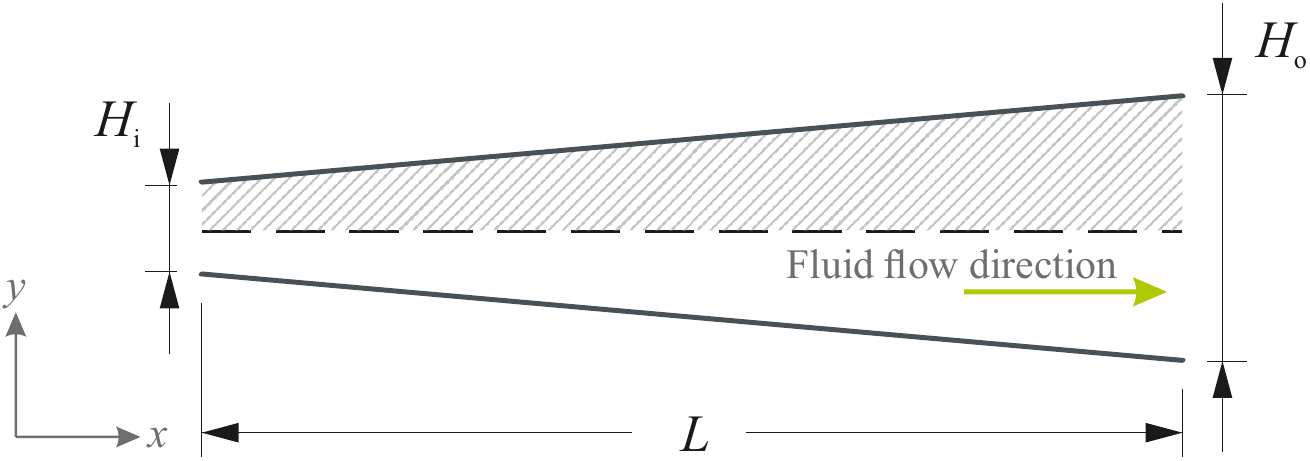}
	\caption{Schematic of the~divergent microchannel considered in the~present study. The~hatched region was considered for calculations.}
	\label{fig:schematic}
\end{figure}

A probabilistic numerical approach based on the~direct simulation Monte-Carlo method is employed to study steady-state, pressure-driven, Poiseuille gaseous nitrogen flow in divergent microchannels shown schematically in Figure~\ref{fig:schematic}. The~divergence angle of the~channel ($\phi$) is defined as follows~\cite{Ebrahimi_2017}:

\begin{equation}
	\phi = \frac{100 \cdot \left(H_\mathrm{o} - H_\mathrm{i}\right)}{L},
	\label{eq:divergent_angle}
\end{equation}

\noindent
where, $H_\mathrm{i}$ and $H_\mathrm{o}$ are the~height of the~channel at the~inlet and the~outlet, respectively. The~length of the~channel ($L$) is twenty times larger than its inlet height (\textit{i.e.} $L = 20\times H_\mathrm{i} = \SI{8e-6}{\meter}$). The~problem is assumed to be two-dimensional, which is a~valid assumption for channels with considerably large depths~\cite{Shah_2018,Zhen_2007}. The~molecular properties of gaseous nitrogen are summarised in Table~\ref{tab: molecular_properties}, where $d_\mathrm{p}$ is molecular diameter, $m_\mathrm{p}$ molecular mass, $\omega$ the~viscosity-temperature index, $T_\mathrm{ref}$ the~reference temperature in the~viscosity-temperature relation, and $\mu_\mathrm{0}$ the~viscosity at the~reference temperature.

Due to the~symmetric configuration of the~channel and the~flow field, one-half of the~physical domain (the hatched region in Figure~\ref{fig:schematic}) is considered for calculations. Heat and fluid flow in the~channel are described in a~Cartesian coordinate system. The~temperature of the~solid walls ($T_\mathrm{w}$) and the~gas at the~inlet ($T_\mathrm{i}$) is constant equal to \SI{300}{\kelvin}. The~Knudsen number is defined based on the~inlet height, \textit{i.e.} $\mathrm{Kn} = \lambda / H_\mathrm{i}$. The~inlet-to-outlet pressure ratio ($\Pi$) is defined as $p_\mathrm{i} / p_\mathrm{o}$ and the~gas pressure at the~channel inlet~($p_\mathrm{i}$) is calculated based on the~Knudsen number and the~ideal gas law. The~effects of tangential momentum accommodation coefficient~($\alpha$) on thermal and fluid flow fields are also studied in the~present work. The~tangential momentum accommodation coefficient ranges between zero (specular reflection) and one (fully-diffused reflections) depending on internal gas molecule structure, the~molecular mass of the~gas, and surface material~\cite{Acharya_2019,Graur_2009}.

\begin{table}[H]
	\centering
	\caption{Molecular properties of gaseous nitrogen.}
	\begin{tabular}{llllll}
		\toprule
		$d_\mathrm{p}$ [\SI{}{\meter}] & $m_\mathrm{p}$ [\SI{}{\kilogram}] & $DOF_\mathrm{rot}$ [\SI{}{-}] & $\omega$ [\SI{}{-}] & $T_\mathrm{ref}$ [\SI{}{\kelvin}] & $\mu_\mathrm{0}$ [\SI{}{\newton\second\per\square\meter}] \\ \midrule
		$\SI{4.17e-10}{}$              & $\SI{4.65e-26}{}$                 & $2$                           & $\SI{7.4e-1}{}$     & $300$                             & $\SI{1.656e-5}{}$                                         \\ \bottomrule
	\end{tabular} 
	\label{tab: molecular_properties}
\end{table}

\section{Numerical procedure}
\label{numerical_procedure}

An extended version of the~dsmcFoam solver~\cite{White_2018}, which has been developed within the~framework of an~open-source solver (OpenFOAM~\cite{OpenFOAM}), was employed in the~present study. This solver has been meticulously validated for a~variety of benchmark cases~\cite{Scanlon_2015,Palharini_2015,White_2018} including subsonic Poiseuille micro-flows. The~variable hard sphere~(VHS) model was selected to treat the~binary inter-molecular collisions. The~energy exchange between both translational and rotational modes was allowed. The~no-time-counter~(NTC)~\cite{Bird_1994} collision sampling model was chosen. Additionally, the~transient adaptive sub-celling~(TAS) scheme~\cite{Su_2010} was utilised to divide collision cells every time-step according to the~number of particles inside each collision cell to have at least two simulator particles in each subcell. This results in the~selection of collision partners that are located within a~relative distance smaller than the~local molecular mean free path.

Oran~\textit{et al.}~\cite{Oran_1998} suggested a~cell size of $\lambda / 3$ to capture the~gradients in flow fields. In~the~case of low-speed micro-flows, larger grids might be employed in the~stream-wise direction due to insignificant flow gradients~\cite{Alexander_1998,Cai_2000,Shen_2003,Sun_2011}. After a~grid independence test, a~grid with cell sizes of $\lambda / 4$ and $\lambda / 2$ in the~normal and the~stream-wise directions, respectively, was selected. To capture the~flow gradients for the~cases with \mbox{$\mathrm{Kn} > \SI{2.5e-1}{}$}, the~grid size was the~same as that for the~cases with $\mathrm{Kn} = \SI{2.5e-1}{}$. All simulations were initialised with $50$ particles per cell to minimise statistical relations between particles~\cite{Shu_2005} and to have at least $15$ particles in the~computational cells adjacent to the~channel’s outlet. The~time-step was selected sufficiently smaller than the~local mean collision time so particles stay inside a~cell for multiple time-steps. The~results of the~grid, time-step size and particle-per-cell independence tests are presented in Figure~\ref{fig:gs_ts_ppc}. To return the~relevant macroscopic data, the~results were sampled over $\SI{3e6}{}$~time-steps after achieving the~steady-state condition. The~sample size was sufficiently large to let the~solution procedure continues even after equating the~inlet and outlet mass flow rates; leading to a~converged solution with negligible (less than $1\%$) statistical noises~\cite{Radtke_2011,Hadjiconstantinou_2003,White_2013}. Each simulation was carried out in parallel on four cores of an~Intel Core i7-3520M processor, and took roughly $\SI{200}{\hour}$ to complete.

\section{Model validation and verification}
\label{sec:model_verification}

\begin{figure}[H] 
	\centering
	\includegraphics[width=0.8\linewidth]{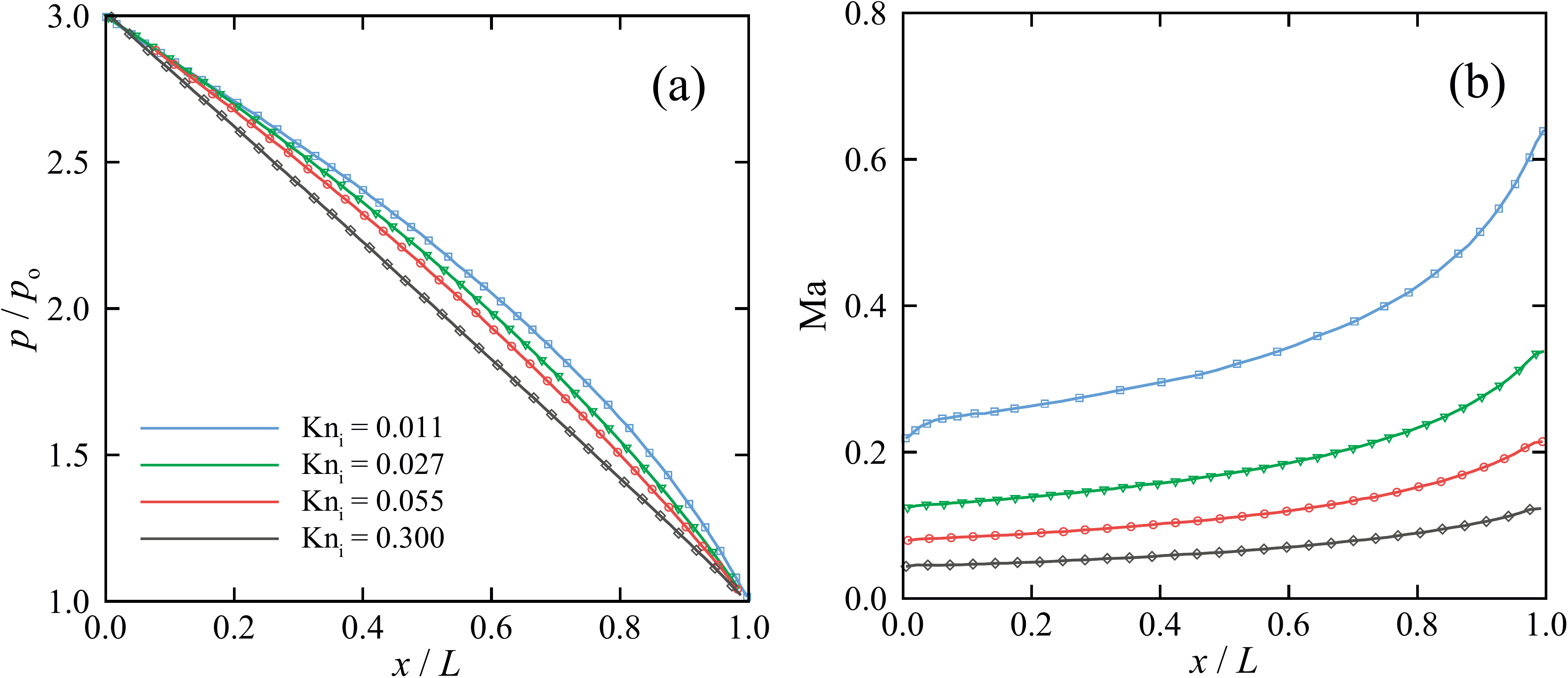}
	\caption{Comparison of the~numerical results obtained from the~present model (solid lines) with the~data reported by \mbox{White~\textit{et al.}~\cite{White_2013}} (symbols) for pressure-driven rarefied argon flow in a~microchannel with uniform cross-section. Profiles of (a) pressure, and (b) Mach number along the~microchannel centreline. }
	\label{fig:validation}
\end{figure}

The Poiseuille argon flow in a~microchannel with a~uniform cross-section was considered to verify the~reliability of the~present model. The~microchannel length is \SI{15}{\micro\meter} and its height is \SI{1}{\micro\meter}. Gaseous argon enters the~channel with a~constant temperature of \SI{300}{\kelvin}, and the~inlet-to-outlet pressure ratio $\Pi$ was set to 3. The~channel walls were assumed to be fully diffuse (\textit{i.e.} $\alpha = 1$), and their temperature was set to \SI{300}{\kelvin}. Four different cases with different Knudsen numbers are studied, which covers the~slip and transition rarefaction regimes. For each case about \SI{1.5e6}{} DSMC simulator particles were utilised. The~predicted pressure and Mach number (Ma) distributions along the~microchannel centreline are shown in Figure~\ref{fig:validation}. The~Mach number is defined as follows:

\begin{equation}
	\mathrm{Ma} = \frac{\lVert \vec{V} \rVert}{\sqrt{\gamma R T}},
	\label{eq:mach_number}
\end{equation}

\noindent
where, $\vec{V}$ is the~fluid velocity vector, $\gamma$ the~ratio of specific heat of nitrogen at a~constant pressure to its specific heat at a~constant volume, $R$~the~specific gas constant, and $T$~the~temperature. The~results are in reasonable agreement with the~data reported by White~\textit{et al.}~\cite{White_2013}. Further comparison of the~results obtained from the~present model with experimental, analytical and numerical data was performed by the~authors for gaseous nitrogen flow in divergent microchannels, and can be found in~\cite{Ebrahimi_2016,Ebrahimi_2017,Varade_2015}.

\begin{figure}[H] 

\includegraphics[width=1.0\linewidth]{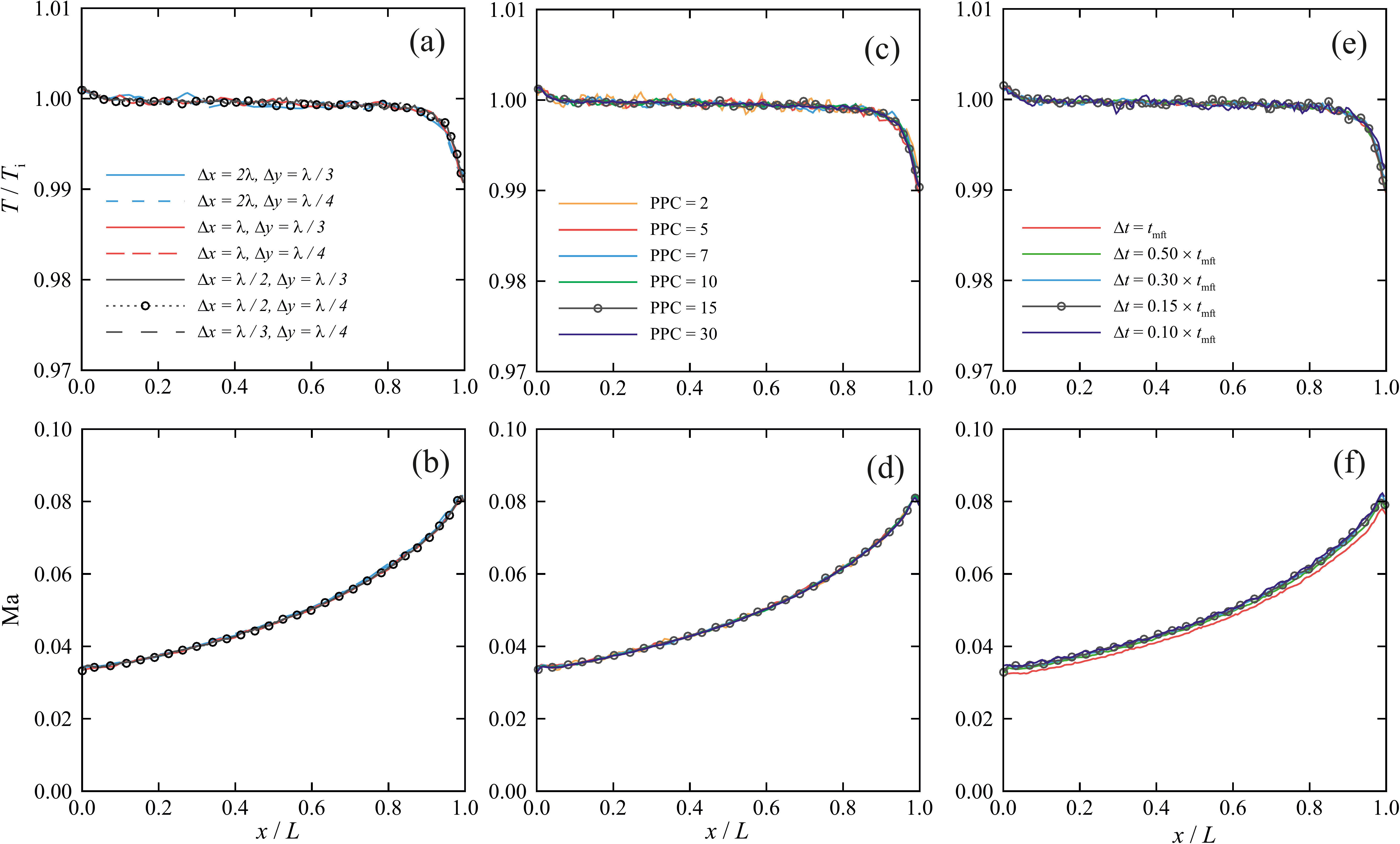}
\caption{The effects of (a and b) the~computational grid size, (c and d) number of simulator particles per cell, and (e and f) time-step size on the~predicted profiles of temperature and Mach number along the~channel centreline. $\mathrm{Kn}_\mathrm{i} = 0.1$, $\phi = 1$ and $\Pi = 2.5$. Curves shown with symbols indicate the~chosen values.}
\label{fig:gs_ts_ppc}
\end{figure}

Figure~\ref{fig:gs_ts_ppc} shows the~influence of grid size, number of simulator particles per cell~(PPC), and time-step size on the~results obtained from the~present DSMC simulations. For this study, nitrogen flow in a~microchannel with uniform cross-section ($\phi = 1$), $\mathrm{Kn}_\mathrm{i} = 0.1$ (transition rarefaction regime), and the~inlet-to-outlet pressure ratio $\Pi$ of 2.5 was considered. To determine an~appropriate time-step size $\Delta t$ within which the~molecular movement and collision are distinguishable, the~following correlations were employed:

\begin{equation}
	\Delta t_\mathrm{c} = \frac{0.2 \lambda}{\sqrt{\frac{2 \kappa_\mathrm{b} T}{m_\mathrm{p}}}},
	\label{eq:time_collision}
\end{equation}

\begin{equation}
	\Delta t_\mathrm{t} = \frac{0.5 \Delta x}{\sqrt{\frac{2 \kappa_\mathrm{b} T}{m_\mathrm{p}}}},
	\label{eq:time_transit}
\end{equation}

\noindent
where, $\Delta t_\mathrm{c}$ is the~mean collision time, $\Delta t_\mathrm{t}$ the~mean transit time and $\kappa_\mathrm{b}$  the~Boltzmann constant. The~time-step size $\Delta t$ was defined as a~fraction of the~minimum value of $\Delta t_\mathrm{c}$ and $\Delta t_\mathrm{t}$ (\textit{i.e.} $\Delta t$ = $\xi \cdot \min\left(\Delta t_\mathrm{c}, \Delta t_\mathrm{t}\right)$ and $\xi \leq 1$). For the~range of parameters studied in the~present work, the~results are practically insensitive to the~grid size, number of simulator particles per cell, and time-step size.

\section{Results and Discussion}
\label{sec:results}
\begin{figure}[H] 
	\centering
	\includegraphics[width=0.8\linewidth]{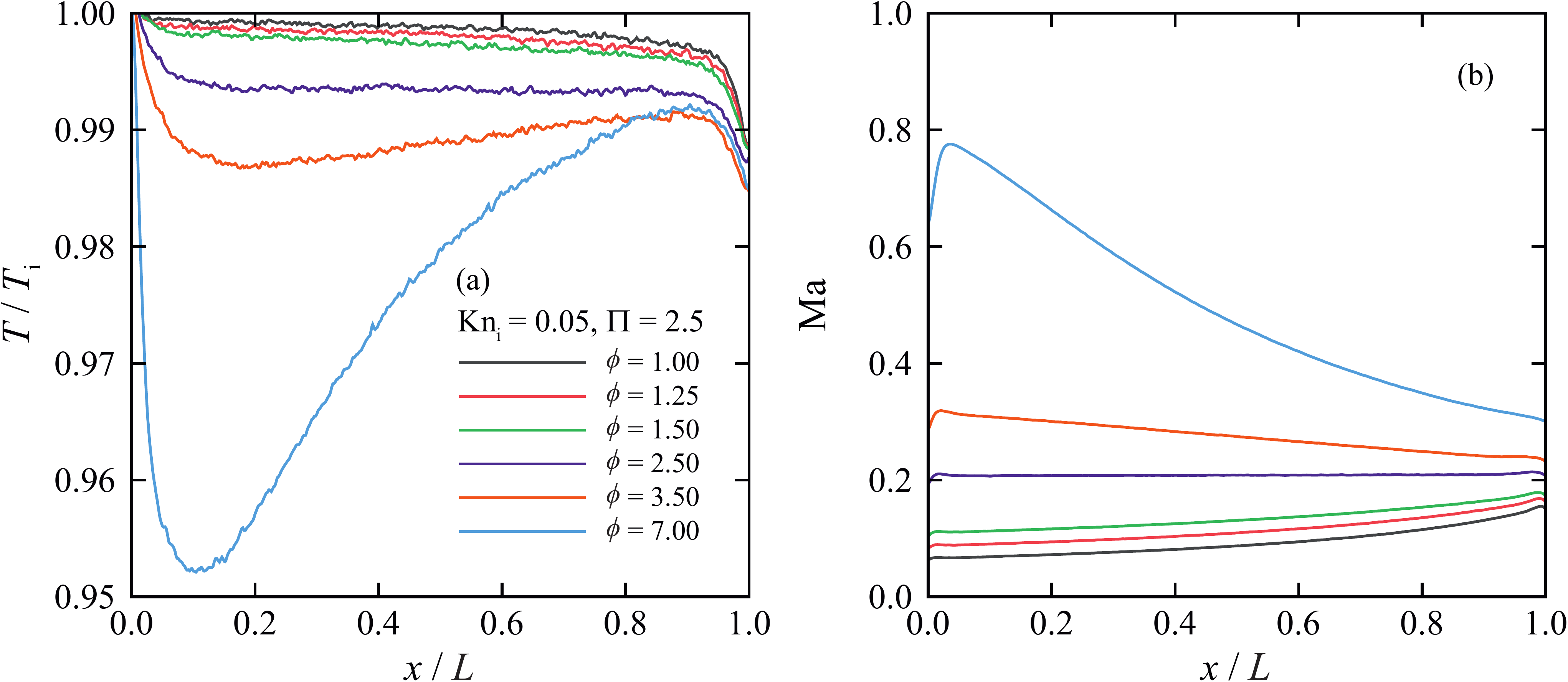}
	\caption{Distribution of (a) gas temperature and (b) Mach number along the~microchannel centreline for different divergence angles $\phi$ ($\mathrm{Kn}_\mathrm{i} = 0.05$ and $\Pi = 2.5$). The~inlet temperature ($T_\mathrm{i}$) is used to non-dimensionalise the~predicted temperatures.}
	\label{fig:temp_mach_centreline_divAng}
\end{figure}

The influence of the~microchannel divergence angle on variations of temperature and Mach number along the~microchannel centreline is shown in Figure~\ref{fig:temp_mach_centreline_divAng}. For \mbox{$\phi < 2.5$}, the~gas temperature decreases along the~channel because of the~augmentation of the~gas velocity towards the~channel outlet. The~increase in kinetic energy of the~gas results in a~reduction of its internal energy and thus its temperature. The~gas velocity remains almost constant along the~channel with $\phi = 2.5$ and the~gas temperature hardly changes along the~channel, except in regions close to the~inlet and outlet boundaries. Further increase in the~divergence angle~$\phi$ results in a~decrease of gas velocity along the~channel, which is attributed to the~increased cross-sectional area of the~channel as well as the~increased shear stress at the~channel walls~\cite{Ebrahimi_2017,Varade_2015}. The~kinetic energy of the~gas converts into internal energy as the~fluid velocity decreases towards the~channel outlet for the~cases with $\phi > 2.5$, leading to an~increase in gas temperature. Additionally, the~bulk gas temperature reduces with an~increase in divergence angle of the~channel because of the~increased bulk gas velocity and the~reduced rate of molecular collision. Because of the~development of a~viscous boundary layer close to the~channel inlet~\cite{Ebrahimi_2017,White_2013}, the~gas flow accelerates in the~entrance region; however, the~flow decelerates after passing this region moving towards the~channel outlet to conform to the~pressure prescribed at the~outlet. Close to the~channel outlet ($x / L > 0.9$), the~gas temperature drops that is attributed to the~rapid gas expansion in that region, which is known as `the~expansion cooling phenomenon'~\cite{Hadjiconstantinou_2001,Zheng_2002}. A~similar behaviour close to the~channel outlet has been reported by others~\cite{Liou_2001,John_2013,Gavasane_2017,Shah_2018,Balaj_2017} for Poiseuille micro-flows in microchannels with uniform cross-sections.

\begin{figure}[H] 
	\centering
	\includegraphics[width=0.8\linewidth]{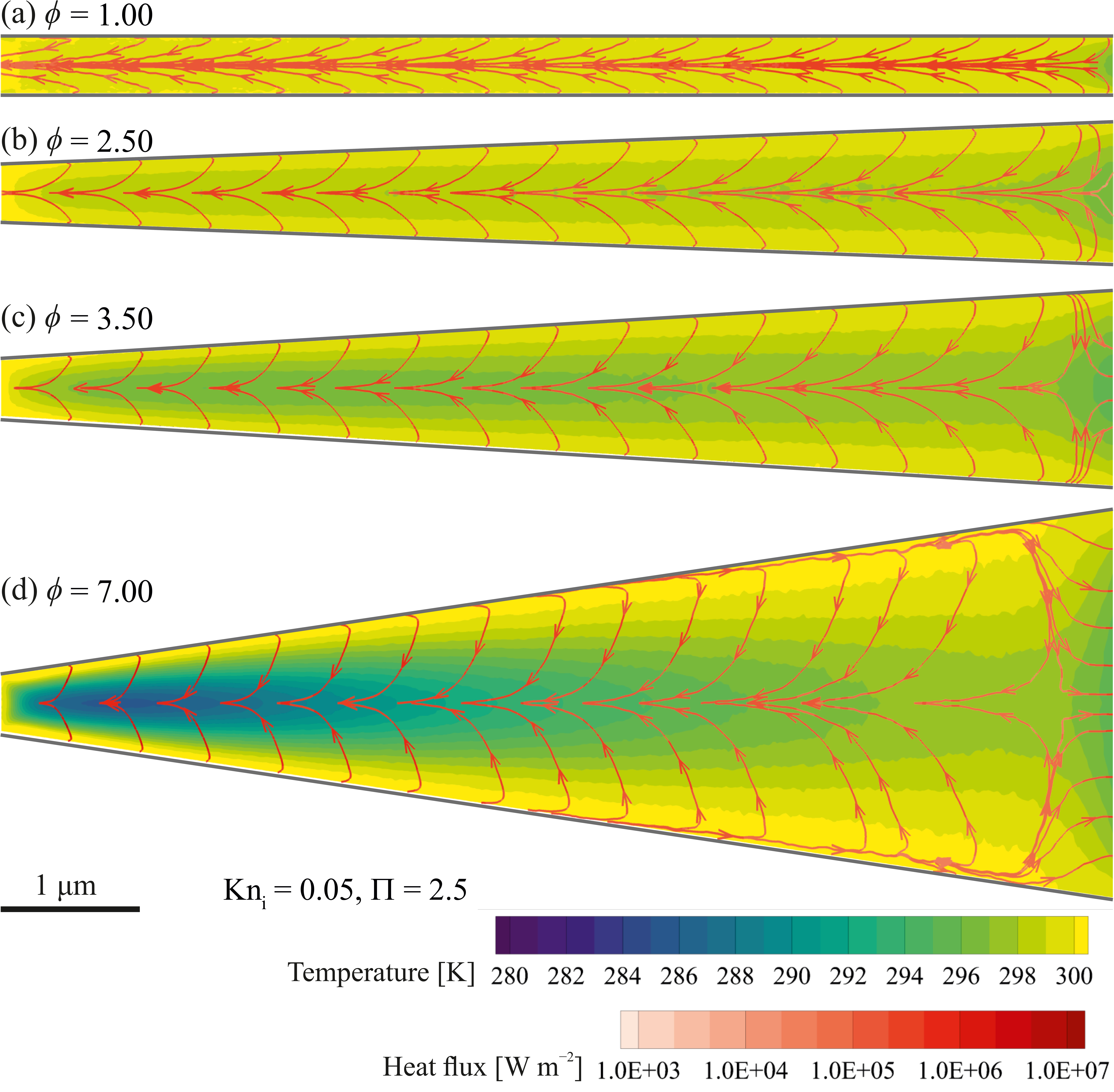}
	\caption{Contours of temperature overlaid with heatlines (coloured by the~magnitude of heat flux) obtained from the~simulations for microchannels with different divergence angles $\phi$. ($\mathrm{Kn}_\mathrm{i} = 0.05$ and $\Pi = 2.5$)}
	\label{fig:heatline_divAng}
\end{figure}

The overall heat flow direction, for the~problem considered in the~present work, is determined by both the~thermal and pressure gradients in the~microchannel~\cite{Sone_2007}. Heat flow direction and contours of dimensionless temperature ($T / T_\mathrm{i}$) obtained from the~DSMC simulations are shown in Figure~\ref{fig:heatline_divAng} for microchannels with different divergence angles at $\mathrm{Kn}_\mathrm{i} = 0.05$ and $\Pi = 2.5$. Heat flows generally towards the~channel inlet except in regions adjacent to the~walls in the~Knudsen layer, where the~heat flow is dominated by viscous heating. For the~cases with $\phi > 2.5$, the~amount of heat induced by viscous dissipation decreases along the~microchannel due to the~reduction of wall shear stress and slip velocity~\cite{Ebrahimi_2017}, decreasing the~influence of viscous dissipation on net heat flow direction in regions close the~walls. In contrast, the~amount of heat generated by viscous dissipation increases as the~gas moves along the~channels with $\phi \leq 2.5$, increasing its influence on net heat flow direction in regions adjacent to the~walls. Accordingly, it can be argued that the~heat flow is dominated by the~pressure gradient (known as a~rarefaction effect~\cite{Sone_2007}) in microchannels with a~divergence angle $\phi \leq 2.5$. Due to the~rarefaction effects, heat flow from cold to hot regions is observed, which is not predictable using the~NSF equations because of the~neglect of high-order rarefaction terms~\cite{Gu_2007}. Heat flow in the~central region of the~channel is dominated by the~mass flow driven by the~pressure gradient. However, the magnitude of thermal gradient increases with an~increase in divergence angle $\phi$, affecting the~net heat flow direction. For the~cases with $\phi > 2.5$, a~thermal separation occurs at certain location in the~channel beyond which heat flows towards the~channel outlet. Thermal separation for pressure-driven gas flow in microchannels of variable cross-sections is attributed to the~enhanced contribution of thermal gradient to the~total heat transfer~\cite{Mahdavi_2015}. The~position of thermal separation point moves towards the~inlet with an~increase in the~divergence angle of the~channel. It should be noted that no hydrodynamic separation was observed for the~cases considered in the~present study~\cite{Ebrahimi_2017,Ebrahimi_2016}. These observations can also be explained through the~asymptotic analysis of the~Boltzmann equation~\cite{Sone_2007} for small Knudsen numbers (\textit{i.e.} slip and early-transition regimes). However, this type of analysis, which is based on the~continuum theory, fails to predict heat transfer accurately at high Knudsen numbers~\cite{Ebrahimi_2016}.

\begin{figure}[H] 
	\centering
	\includegraphics[width=0.8\linewidth]{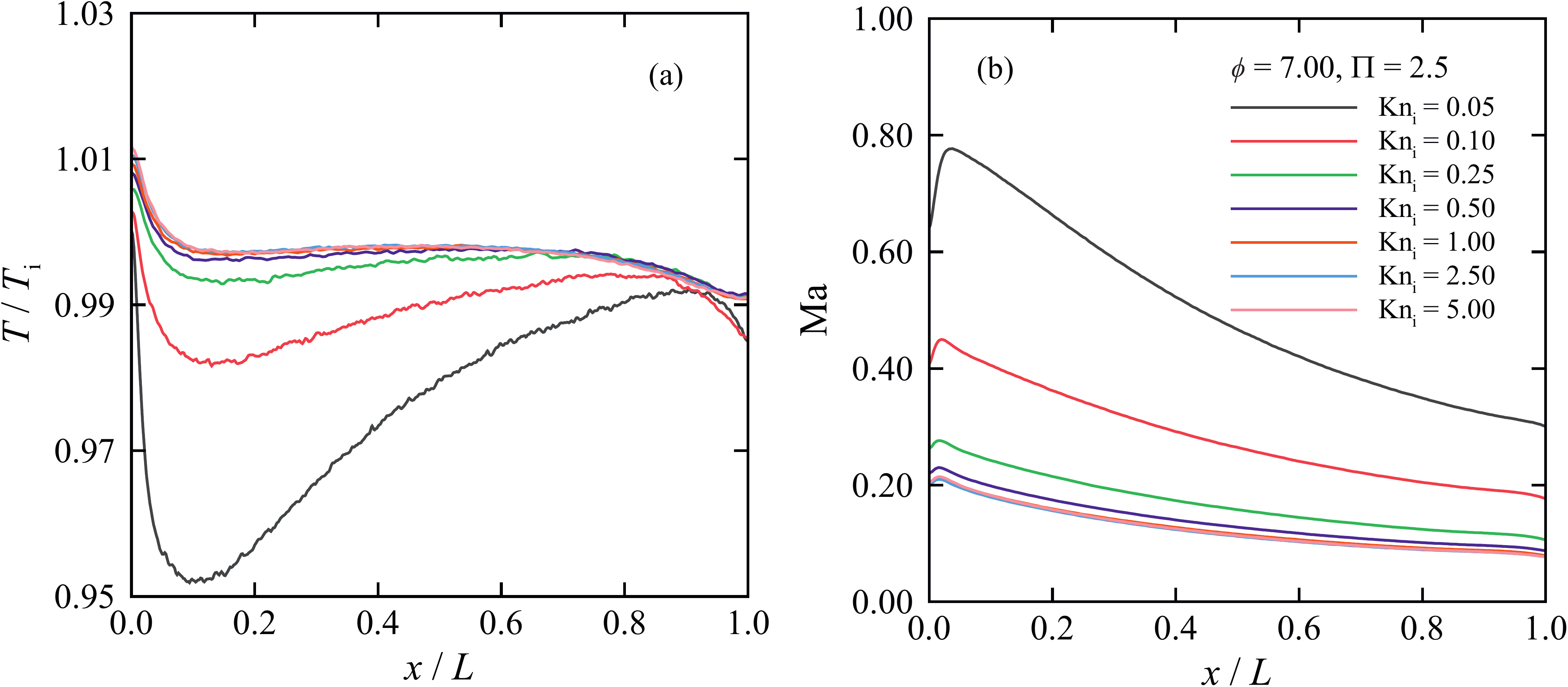}
	\caption{The effect of rarefaction on the~profiles of (a) gas temperature and (b) Mach number along the~microchannel centreline ($\phi = 7.00$ and $\Pi = 2.5$). The~inlet temperature ($T_\mathrm{i}$) is used to non-dimensionalise the~predicted temperatures.}
	\label{fig:temp_mach_centreline_kn}
\end{figure}

Numerical predictions of the~gas temperature and Mach number along the~channel centreline are presented in Figure~\ref{fig:temp_mach_centreline_kn} for different inlet Knudsen numbers~$\mathrm{Kn}_\mathrm{i}$ at $\phi = 7$ and $\Pi = 2.5$. The~bulk gas velocity along the~channel and variations in the~gas velocity decreases with an~increase in the~Knudsen number. However, this reduction in the~velocity magnitude and its variations becomes insignificant for $\mathrm{Kn}_\mathrm{i} \geq 1$. Variation of gas velocity along the~channel and the~molecular collision frequency decreases with an~increase in Knudsen number, limiting the~variation of gas temperature along the~channel. The~magnitude of temperature drop close to the~channel outlet also decreases with an increase in the~inlet Knudsen number. Moreover, the~predicted gas temperatures close to the~channel inlet are slightly higher than the~inlet gas temperature $T_\mathrm{i}$, which is because of the~augmentation of temperature jump with increasing the~Knudsen number~\cite{John_2013}.

\begin{figure}[H] 

\includegraphics[width=1.0\linewidth]{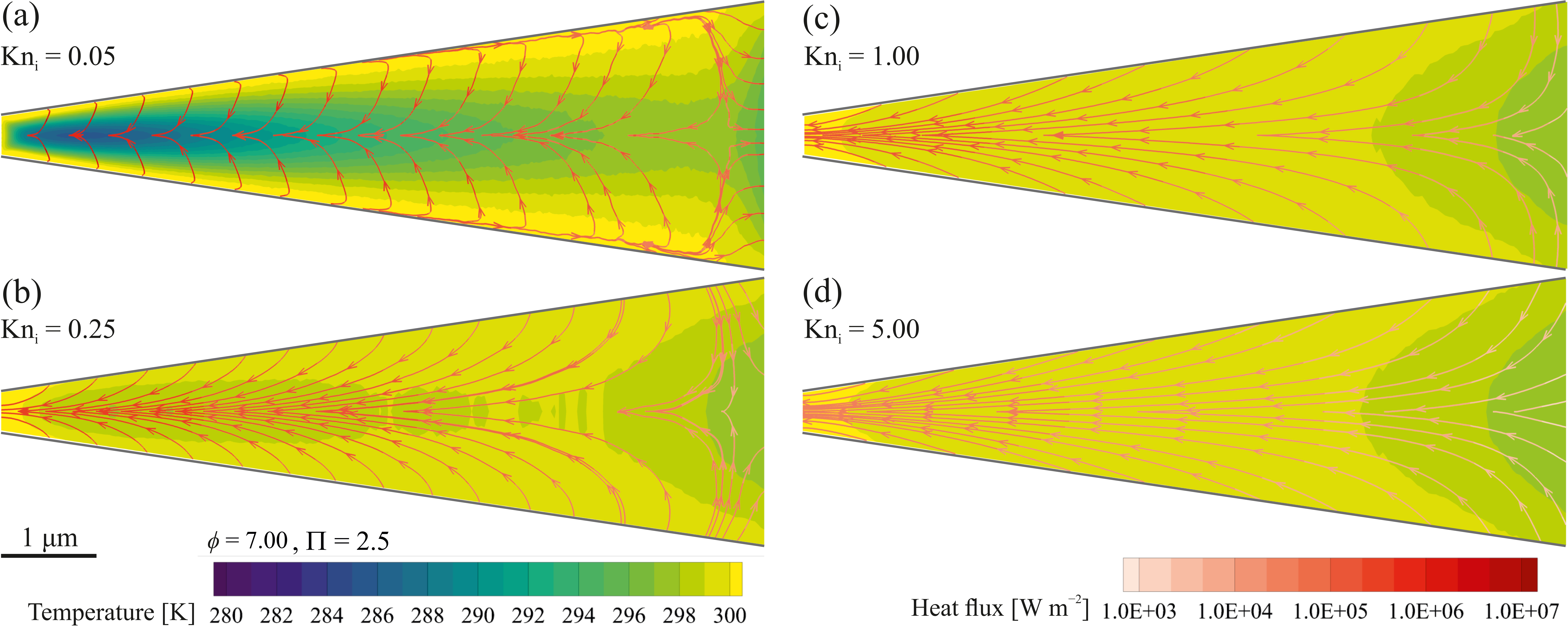}
\caption{Heatlines (coloured by the~magnitude of heat flux) overlaid on contours of dimensionless temperature ($T / T_\mathrm{i}$) predicted for different inlet Knudsen numbers $\mathrm{Kn}_\mathrm{i}$. $\phi = 7.00$ and $\Pi = 2.5$.}
\label{fig:heatline_kn}
\end{figure}

The influence of the~Knudsen number on the~thermal field and the~heat flow direction in a~divergent channel with $\phi = 7$ and $\Pi = 2.5$ is shown in Figure~\ref{fig:heatline_kn}. The~temperature gradient in the~channel decreases with an~increase in Knudsen number, decreasing the~contribution of the~Fourier heat transfer in net heat flow. According to Figure~\ref{fig:heatline_kn} (c) and (d), at sufficiently large $\mathrm{Kn}_\mathrm{i}$ the~heat flow in the~channel is merely towards the~channel inlet that demonstrates the~domination of the~pressure gradient in the~net heat flow direction, intensifying the~counter-gradient cold-to-hot heat transfer (also known as anti-Fourier heat transfer). Moreover, the~total heat flow rate decreases with increasing the~Knudsen number. The~distance between the~thermal separation point and the~channel outlet increases with a~decrease in $\mathrm{Kn}_\mathrm{i}$.

\begin{figure}[H] 
	\centering
	\includegraphics[width=0.8\linewidth]{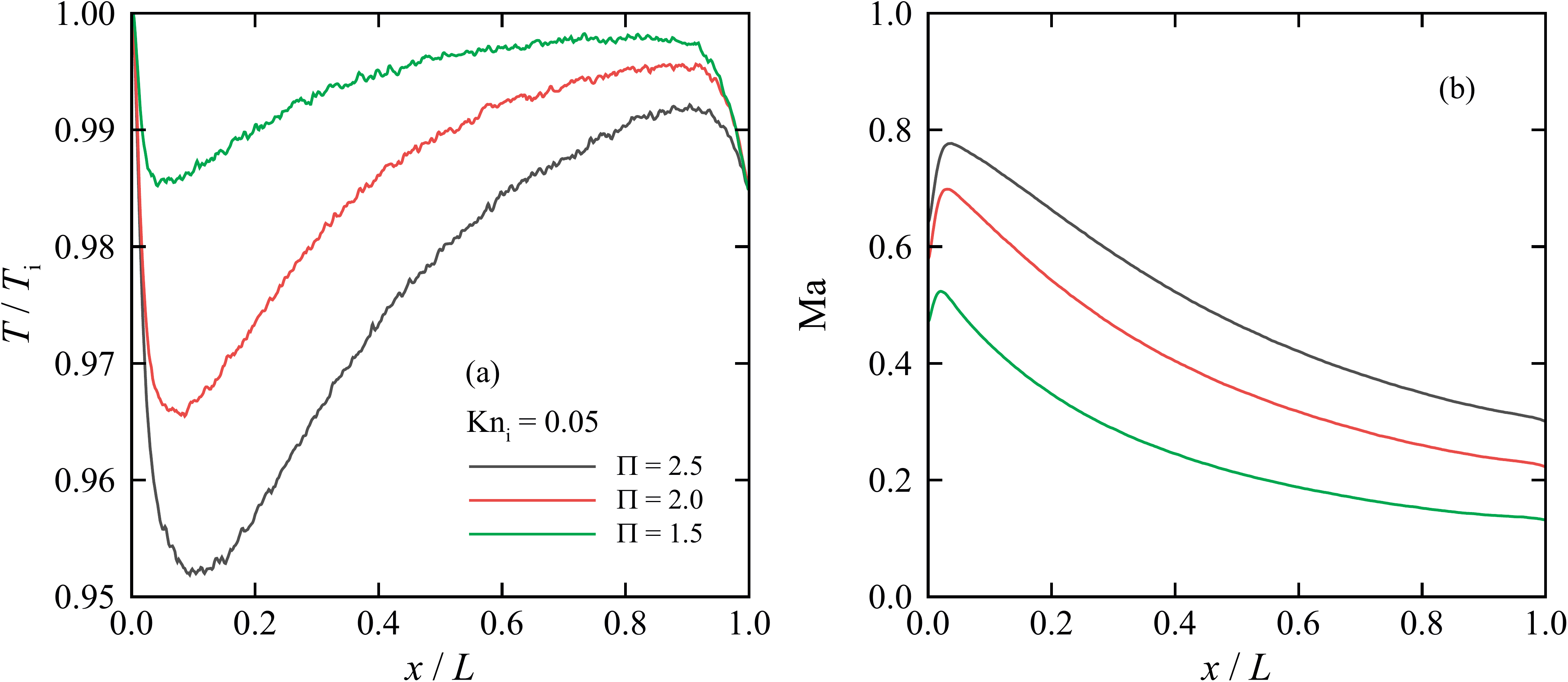}
	\caption{The effect of the~inlet-to-outlet pressure ratio $\Pi$ on the~profiles of (a) gas temperature and (b) Mach number along the~microchannel centreline ($\phi = 7.00$ and $\mathrm{Kn} = 0.05$). The~inlet temperature ($T_\mathrm{i}$) is used to non-dimensionalise the~predicted temperatures.}
	\label{fig:temp_mach_centreline_pr}
\end{figure}

\begin{figure}[H] 
\centering

\includegraphics[width=1.0\linewidth]{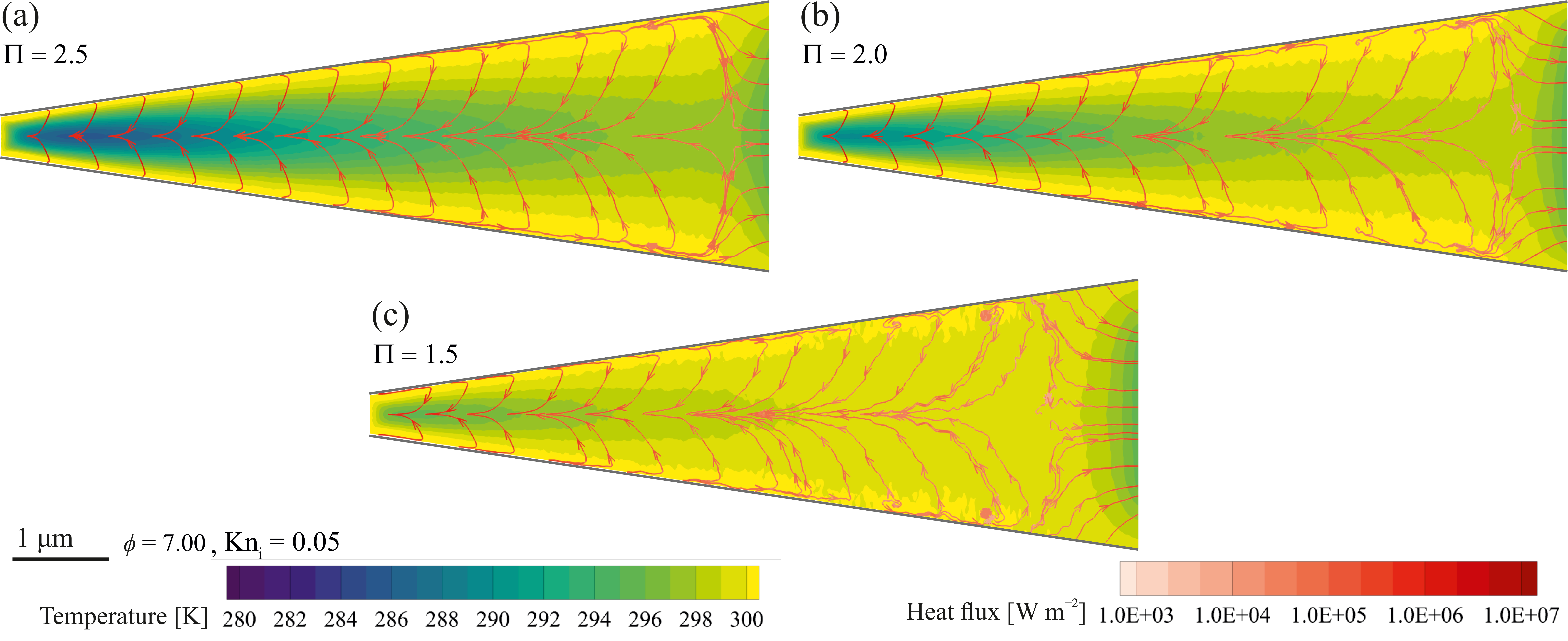}
\caption{The influence of the~inlet-to-outlet pressure ratio $\Pi$ on thermal field (contours) and heat flow direction (coloured by the~magnitude of heat flux). $\mathrm{Kn}_\mathrm{i} = 0.05$ and $\phi = 7$.}
\label{fig:heatline_pr}
\end{figure}

The~bulk gas velocity and temperature in the~channel decrease with a~decrease in the~inlet-to-outlet pressure ratio~$\Pi$, as shown in Figure~\ref{fig:temp_mach_centreline_pr}. Increasing the~inlet-to-outlet pressure ratio increases the~gas kinetic energy and decreases its internal energy. The~thermal field as well as the~heat lines are shown in Figure~\ref{fig:heatline_pr} for channels with a~divergence angle of $\phi = 7$ and Knudsen number of 0.05. The~magnitude of thermal gradient close to the~channel outlet increases with decreasing the~pressure ratio~$\Pi$, which enhances the~contribution of the~thermal gradient to the~net heat flow; this affects the~net heat flow direction and leads to the~occurrence of thermal separation farther away from the~channel outlet. The~heat generated in the~Knudsen layer due to viscous dissipation diminishes with reducing the~inlet-to-outlet pressure ratio~$\Pi$, decreasing the~contribution of thermal gradients to the~net heat flow close to the~channel walls. 

\begin{figure}[H] 
	\centering
	\includegraphics[width=0.8\linewidth]{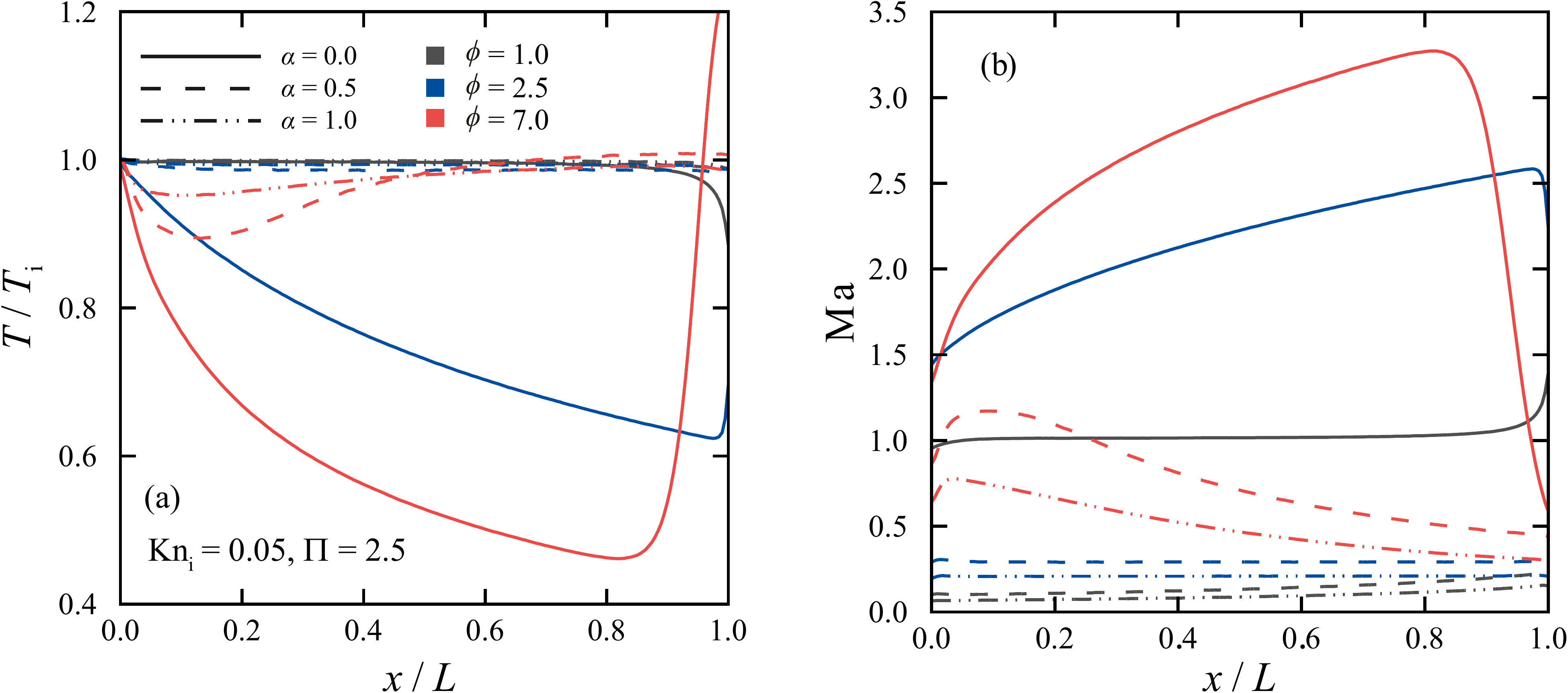}
	\caption{The effect of the~tangential accommodation coefficient $\alpha$ on the~profiles of (a) gas temperature and (b) Mach number along the~microchannel centreline ($\Pi = 2.5$ and $\mathrm{Kn}_\mathrm{i} = 0.05$). The~inlet temperature ($T_\mathrm{i}$) is used to non-dimensionalise the~predicted temperatures.}
	\label{fig:temp_mach_centreline_tac}
\end{figure}

The~influence of the~tangential accommodation coefficient ($\alpha$) on the~profiles of temperature and velocity along divergent channels is shown in~Figure~\ref{fig:temp_mach_centreline_tac} for $\Pi = 2.5$ and $\mathrm{Kn}_\mathrm{i} = 0.05$. For all the cases considered in the~present work, decreasing the~tangential accommodation coefficient weakens the~interactions between the~gas molecules and the~channel walls, leading to an~increase in the~gas velocity. A~similar behaviour has also been observed both experimentally and numerically for rarefied gas flows in microchannels with uniform cross-sections~\cite{John_2013,Graur_2009,Acharya_2019,Cercignani_2004}. As expected, the~augmentation of gas velocity in the~channel results in an increase in the~kinetic energy and a~decrease in internal energy and thus the~gas temperature. The~results presented in~Figure~\ref{fig:temp_mach_centreline_tac} show that the~gas reaches supersonic velocities when $\alpha = 0$ (\textit{i.e.}~specular reflection); this happens for the~case with $\phi = 7.0$ only when $\alpha = 0.5$ and does not happen when $\alpha = 1$ (\textit{i.e.}~fully-diffused reflection). 

\begin{figure}[H] 
\centering

\includegraphics[width=1.0\linewidth]{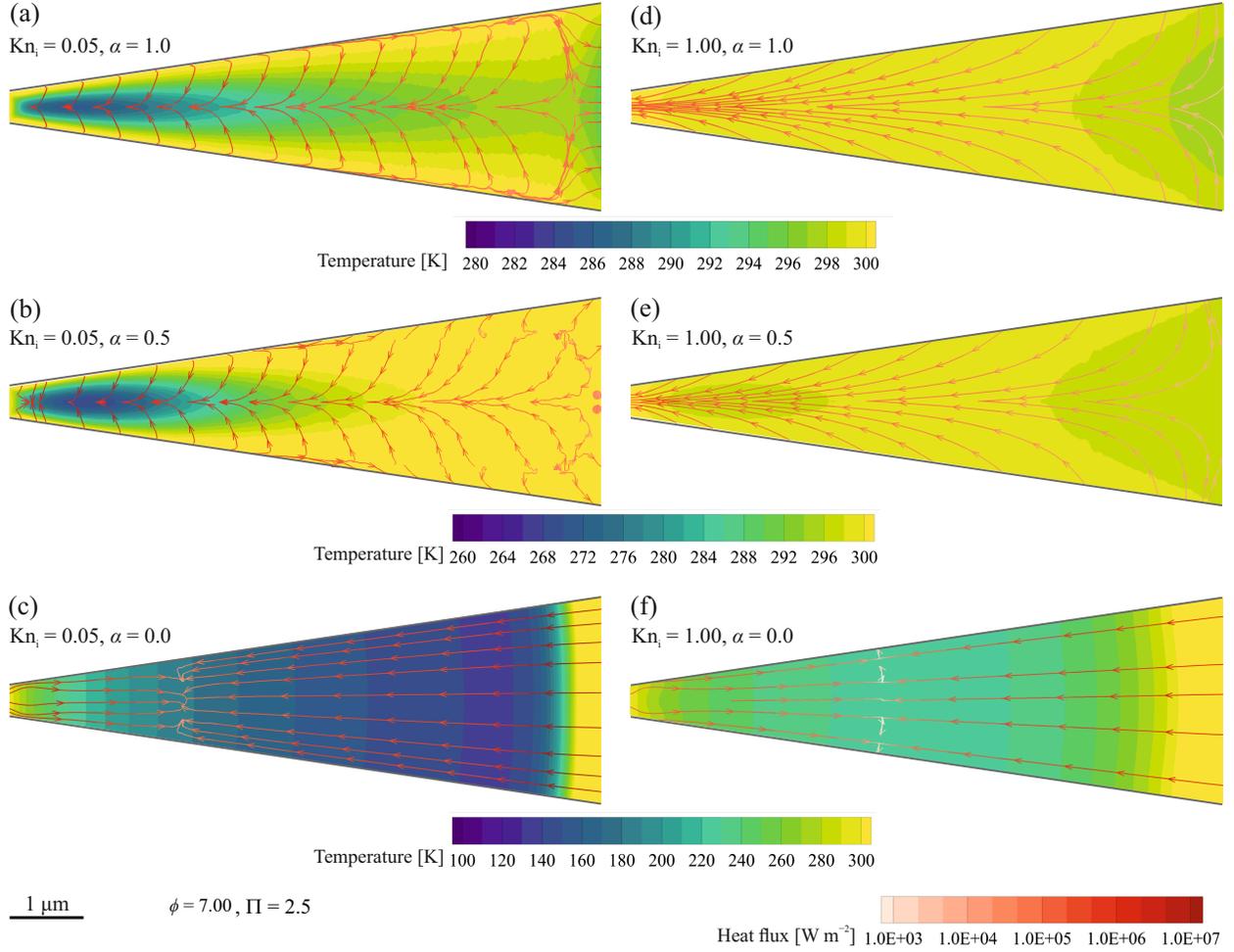}
\caption{The influence of the~tangential accommodation coefficient $\alpha$ on thermal field (contours) and heat flow direction (coloured by the~magnitude of heat flux). $\Pi = 2.5$ and $\phi = 7$.}
\label{fig:heatline_tac}
\end{figure}

The~results presented in Figure~\ref{fig:temp_mach_centreline_tac} show that heat and fluid flow patterns in the~channel are sensitive to the~value of the~tangential accommodation coefficient $\alpha$, particularly for low values of the~Knudsen numbers. The~number of gas molecules contributing to the~adiabatic exchange of energy increases with decreasing the~value of the~tangential accommodation coefficient $\alpha$, increasing the~contribution of thermal gradients to the~net heat flow. Reducing the~value of the~tangential accommodation coefficient to $\alpha = 0$ as an~extreme condition, it is observed that the~heat flow converges to a~sink point, where pressure gradients balance thermal gradients. Heat lines are directed towards the~channel outlet in regions where thermal gradients dominate the~net heat flow.

\section{Conclusions}

Two-dimensional numerical simulations based on the~direct simulation Monte-Carlo~(DSMC) method were performed to predict thermal and flow fields of pressure-driven nitrogen flow in divergent microchannels. A~wide range of Knudsen numbers was studied covering the~slip to the~free molecular rarefaction regimes. Moreover, the~influence of microchannel divergence angle, inlet-to-outlet pressure ratio and incomplete surface accommodation were investigated. The~following conclusions are drawn based on the~present study.

the~influence of thermal gradient on net heat flow direction enhances with an~increase in the~divergence angle of the~microchannel and can lead to the~occurrence of thermal separation even when hydrodynamic separation is absent. The~contribution of the~pressure gradient in determining the~heat flow direction enhances with an~increase in the~Knudsen number. At sufficiently large Knudsen numbers, the~heat flow is merely towards the~channel inlet. Cold-to-hot heat transfer (also known as anti-Fourier heat transfer) was observed in the~microchannel as a~result of nonequilibrium effects. Viscous dissipation in the~Knudsen layer diminishes with a~decrease in applied pressure ratio, which results in the~domination of the~pressure gradient influence on heat flow close to the~channel walls.

Further studies may focus on realising the effects of the molecular weight and structure, or different gas mixtures on thermal and fluid flow fields inside micro- and Nano-channels with with complex geometries.

\section*{Author Contributions}
\label{sec:author_contributions}

Conceptualization, A.E. and E.R.; methodology, A.E.; software, A.E. and V.S.; validation, A.E.; formal analysis, A.E.; investigation, A.E.; resources, A.E., V.S. and E.R; data curation, A.E. and V.S.; writing---original draft preparation, A.E.; writing---review and editing, A.E., V.S. and E.R.; visualization, A.E. and V.S.; supervision, A.E. and E.R.; project administration, A.E. All authors have read and agreed to the published version of the manuscript.

\section*{Conflict of interest}
\label{sec:conflict_of_interest}

The authors declare no conflict of interest.

\section*{Data availability}
\label{sec:data_availability}

The data presented in this study are available on request from the corresponding author.

\small{
	\bibliographystyle{elsarticle-num}
	\bibliography{ref}
}

\end{document}